\documentclass[preprintnumbers, prd, twocolumn, showpacs, floatfix,
preprintnumbers, letterpaper, superscriptaddress]{revtex4}
\usepackage[dvips]{graphicx}
\usepackage{epsf}
\usepackage{amsmath}
\usepackage{amssymb}
\usepackage{graphicx}
\usepackage{dcolumn}
\usepackage{bm}
\usepackage{color}

\newcommand {\lla} {\ {\raise-.5ex\hbox{$\buildrel<\over\sim$}}\
}

\def\be{\begin{equation}}
\def\ee{\end{equation}}
\def\bea{\begin{eqnarray}}
\def\eea{\end{eqnarray}}

\newcommand{\tc}{\textcolor{black}}

\begin{document}

\title{Matter Bounce Cosmology with the $f(T)$ Gravity}

\author{Yi-Fu Cai}
\email{ycai21@asu.edu}
\affiliation{Department of Physics, Arizona State University, Tempe, AZ
85287, USA}
\author{Shih-Hung Chen}
\email{schen102@asu.edu}
\affiliation{Department of Physics, Arizona State University, Tempe, AZ
85287, USA}
\author{James~B.~Dent}
\email{jbdent@asu.edu}
\affiliation{Department of Physics, Arizona State University, Tempe, AZ
85287, USA}
\author{Sourish Dutta}
\email{sourish.d@gmail.com}
\affiliation{Department of Physics, Arizona State University,
Tempe, AZ 85287, USA}
\author{Emmanuel N. Saridakis}
\email{Emmanuel_Saridakis@baylor.edu}
\affiliation{CASPER, Physics Department, Baylor University, Waco, TX
76798-7310, USA}
\affiliation{National Center for Theoretical Sciences, Hsinchu, Taiwan 300}


\begin{abstract}
We show that the $f(T)$ gravitational paradigm, in which gravity is
described by an arbitrary function of the torsion scalar, can provide a
mechanism for realizing bouncing cosmologies, thereby avoiding the Big Bang
singularity. After constructing the simplest version of an $f(T)$ matter
bounce, we investigate the scalar and tensor modes of cosmological
perturbations. Our results show that metric perturbations in the scalar
sector lead to a background-dependent sound speed, which is a
distinguishable feature from Einstein gravity. Additionally, we obtain a
scale-invariant primordial power spectrum, which is consistent with
cosmological observations, but suffers from the problem of a large
tensor-to-scalar ratio. However, this can be avoided by introducing extra
fields, such as a matter bounce curvaton.
\end{abstract}

\maketitle


\section{Introduction}

Inflation is now considered to be a crucial part of the cosmological
history of the universe  \cite{inflation}, however the so called ``standard
model of the universe'' still faces the problem of the initial singularity.
Such a singularity is unavoidable if inflation is realized using a scalar
field while the background spacetime is described by the standard Einstein
action \cite{Borde:1993xh}. As a consequence, there has been much energy
expended towards resolving this problem, e.g. through null-energy-condition
violating quantum fluctuations (``island cosmology models''  \cite{dutta1,
dutta2}), quantum gravity effects, or effective field theory techniques.

A potential solution to the cosmological singularity problem may be
provided by non-singular bouncing cosmologies \cite{Mukhanov:1991zn}. Such
scenarios have been constructed through various approaches to modified
gravity, such as the Pre-Big-Bang \cite{Veneziano:1991ek} and the Ekpyrotic
\cite{Khoury:2001wf} models, gravitational actions with higher order
corrections \cite{Brustein:1997cv}, braneworld scenarios
\cite{Kehagias:1999vr}, non-relativistic gravity
\cite{Brandenberger:2009yt, Cai:2009in}, loop quantum cosmology
\cite{Bojowald:2001xe} or in the frame of a closed universe
\cite{Martin:2003sf}. {{For a review on models of modifying higher
derivative gravity, which solve cosmic singularities in an efficient way,
we refer to Ref.
\cite{Nojiri:2010wj}.}} Non-singular bounces may be alternatively
investigated using effective field theory techniques, introducing matter
fields which violate the null energy condition \cite{Cai:2007qw,
Cai:2009zp} or introducing non-conventional mixing terms
\cite{Saridakis:2009jq}. The extension of all the above bouncing scenarios
is the (old) paradigm of cyclic cosmology \cite{tolman}, in which the
universe experiences a periodic sequence of contractions and expansions,
which has been reawakened recently  \cite{Steinhardt:2001st}, since it
brings different insights to the origin of the observable universe
\cite{Xiong:2008ic, Lidsey:2004ef, cyclic, Cai:2010zma} (see
\cite{Novello:2008ra} for a review).

Along separate lines, $f(T)$ gravity has recently received attention in the
literature  \cite{Ferraro:2006jd,
Bengochea:2008gz, Linder:2010py}, mostly in the context of explaining the
observed acceleration of the Universe. It is based on the old idea of the
``teleparallel" equivalent of General Relativity (TEGR) \cite{ein28,
Hayashi79}, which, instead of using the curvature defined via the
Levi-Civita connection, uses the Weitzenb{\"{o}}ck connection that has no
curvature but only torsion. The dynamical objects in such a framework are
the four linearly independent vierbeins. The advantage of this framework is
that the torsion tensor is formed solely from products of first derivatives
of the tetrad. As described in \cite{Hayashi79}, the Lagrangian density,
$T$, can then be constructed from this torsion tensor under the assumptions
of invariance under general coordinate transformations, global Lorentz
transformations, and the parity operation, along with requiring the
Lagrangian density to be second order in the torsion tensor. However,
instead of using the tensor scalar $T$ the authors of
\cite{Bengochea:2008gz,Linder:2010py} generalized the above formalism to a
modified $f(T)$ version, thus making the Lagrangian density a function of
$T$, similar to the well-known extension of $f(R)$ Einstein-Hilbert action.
In comparison with $f(R)$ gravity, whose fourth-order equations may lead to
pathologies, $f(T)$ gravity has the significant advantage of possessing
second-order field equations. This feature has led to
a rapidly increasing interest in the literature, and apart from obtaining
acceleration \cite{Bengochea:2008gz, Linder:2010py} one can reconstruct a
variety of cosmological evolutions \cite{Myrzakulov:2010vz} and solutions
\cite{Wang:2011xf}, add a scalar field \cite{Yerzhanov:2010vu}, use
observational data in order to constrain the model parameters
\cite{Wu:2010mn}, examine the dynamical behavior of the scenario
\cite{Wu:2010xk}, and proceed beyond the background evolution,
investigating the vacuum and matter perturbations \cite{Chen:2010va,
Dent:2011zz} as well as the large-scale structure \cite{Li2011}. Note
however that there is a discussion whether forms of $f(T)$ other than
linear should be expected \cite{Deliduman:2011ga}.

{{One interesting feature of the $f(T)$ theory is that the null energy
condition could be effectively violated. Accompanied with this feature, one
expects to obtain a series of nontrivial phenomenological solutions.
Particularly, it was observed that the violation of null energy condition
is related to the nonsingular bouncing solution in the early universe
\cite{Cai:2007qw}. Therefore, the singularity avoidance can be obtained in
general in $f(T)$ gravity.}} In the present work we are interested in
searching for a scenario of bouncing cosmology in the early universe, which
is governed
by $f(T)$ gravity. As we show, the realization of a big bounce and the
avoidance of singularities is straightforward. Additionally, we investigate
in detail the evolution of perturbations in a specific model of the matter
bounce. Our example illustrates that it is possible to generate a
scale-invariant primordial power spectrum and pass through the nonsingular
bouncing point in the context of $f(T)$ bounce cosmology.

This paper is organized as follows. In section \ref{model} we briefly
review the basic idea of $f(T)$ gravity. In section \ref{matter bounce
background} we investigate the realization of bouncing cosmology by virtue
of $f(T)$ gravity. Specifically, we postulate an explicit form for the
background scale factor, and following this ansatz we reconstruct the
corresponding form of $f(T)$ both analytically and numerically. In section
\ref{perturbation} we study the evolution of cosmological perturbations of
scalar and tensor types along with the matter bounce scenario. Finally,
section \ref{conclusion} is devoted to the summary of the results.

\section{$f(T)$ gravity and cosmology}

\label{model}

In this section we briefly review $f(T)$ gravity and we provide
the
background cosmological equations in a universe governed by such
a modified
gravitational sector. Throughout the work we consider a flat
Friedmann-Robertson-Walker (FRW) background geometry with metric
\begin{equation}
ds^2= dt^2-a^2(t)\,\delta_{ij} dx^i dx^j,
\end{equation}
where $t$ is the cosmic time, $x^i$ are the comoving spatial
coordinates,
and $a(t)$ is the scale factor. In this manuscript our notation
is as
follows: Greek indices $\mu, \nu,$... run over all coordinate
space-time 0,
1, 2, 3, lower case Latin indices (from the middle of the
alphabet) $i, j,
...$ run over spatial coordinates 1, 2, 3, capital Latin indices
$A, B, $...
run over the tangent space-time 0, 1, 2, 3, and lower case Latin
indices
(from the beginning of the alphabet) $a,b, $... will run over the
tangent
space spatial coordinates 1, 2, 3.

\subsection{$f(T)$ gravity}

\label{fTgrav}

As stated in the Introduction, the dynamical variable of the old
``teleparallel'' gravity, as well as its $f(T)$ extension, is the
vierbein
field ${\mathbf{e}_A(x^\mu)}$. This forms an orthonormal basis
for the tangent
space at each point $x^\mu$ of the manifold, that is $\mathbf{e}
_A\cdot%
\mathbf{e}_B=\eta_{AB}$, where $\eta_{AB}=diag (1,-1,-1,-1)$.
Furthermore,
the vector $\mathbf{e}_A$ can be analyzed with the use of its
components $%
e_A^\mu$ in a coordinate basis, that is
$\mathbf{e}_A=e^\mu_A\partial_\mu $.

In such a construction, the metric tensor is obtained from the
dual vierbein
as
\begin{equation}  \label{metrdef}
g_{\mu\nu}(x)=\eta_{AB}\, e^A_\mu (x)\, e^B_\nu (x).
\end{equation}
Contrary to General Relativity, which uses the torsionless
Levi-Civita
connection, in the present formalism ones uses the curvatureless
Weitzenb%
\"{o}ck connection  \cite{Weitzenb23}, whose torsion tensor reads
\begin{equation}  \label{torsion2}
{T}^\lambda_{\:\mu\nu}=\overset{\mathbf{w}}{\Gamma}^\lambda_{
\nu\mu}-%
\overset{\mathbf{w}}{\Gamma}^\lambda_{\mu\nu}
=e^\lambda_A\:(\partial_\mu
e^A_\nu-\partial_\nu e^A_\mu).
\end{equation}
Moreover, the contorsion tensor, which equals the difference
between the Weitzenb%
\"{o}ck and Levi-Civita connections, is defined as
\begin{equation}  \label{cotorsion}
K^{\mu\nu}_{\:\:\:\:\rho}=-\frac{1}{2}\Big(T^{\mu\nu}_{
\:\:\:\:\rho}
-T^{\nu\mu}_{\:\:\:\:\rho}-T_{\rho}^{\:\:\:\:\mu\nu}\Big).
\end{equation}
Finally, it proves useful to define
\begin{equation}  \label{Stensor}
S_\rho^{\:\:\:\mu\nu}=\frac{1}{2}\Big(K^{\mu\nu}_{\:\:\:\:\rho}
+\delta^\mu_\rho
\:T^{\alpha\nu}_{\:\:\:\:\alpha}-\delta^\nu_\rho\:
T^{\alpha\mu}_{\:\:\:\:\alpha}\Big).
\end{equation}
Using these quantities one can define the so called
``teleparallel
Lagrangian'', which is nothing other than the torsion scalar, as
\cite%
{Hayashi79,Maluf:1994ji,Arcos:2005ec}
\begin{equation}  \label{telelag}
T\equiv S_\rho^{\:\:\:\mu\nu}\:T^\rho_{\:\:\:\mu\nu}.
\end{equation}
In summary, in the present formalism all the information
concerning the
gravitational field is included in the torsion tensor ${T}%
^\lambda_{\:\mu\nu} $, and the torsion scalar $T$ arises from it
in a
similar way as the curvature scalar arises from the curvature
(Riemann)
tensor. Finally, the torsion scalar gives rise to the dynamical
equations
for the vierbein, which imply the Einstein equations for the
metric.

While in teleparallel gravity the action is constructed by the
teleparallel
Lagrangian $T$, the idea of $f(T)$ gravity is to generalize $T$
to a
function $T+f(T)$, which is similar in spirit to the
generalization of the
Ricci scalar $R$ in the Einstein-Hilbert action to a function
$f(R)$. In
particular, the action in a universe governed by $f(T)$ gravity
reads:
\begin{eqnarray}  \label{action}
I = \frac{1}{16\pi G}\int d^4x e \left[T+f(T)+L_m\right],
\end{eqnarray}
where $e = \text{det}(e_{\mu}^A) = \sqrt{-g}$ and $L_m$ stands
for the
matter Lagrangian. We mention here that since the Ricci scalar
$R$ and the torsion scalar $T$
differ only by a total derivative  \cite{Weinberg:2008}, in the
case where $%
f(T)$ is a constant (which will play the role of a cosmological
constant)
the action (\ref{action}) is equivalent to General Relativity
with a
cosmological constant.

Lastly, we stress that throughout this work we use the common
choice for the
form of the vierbien, namely
\begin{equation}  \label{weproudlyuse}
e_{\mu}^A=\mathrm{diag}(1,a,a,a).
\end{equation}
It can be easily found that the family of vierbiens related to
(\ref%
{weproudlyuse}) through global Lorentz transformations, lead to
the same
equations of motion. Note however that, as was shown in
\cite{Li:2010cg},
$f(T)$ gravity does not preserve local Lorentz invariance. Thus,
one should
in principle study the cosmological consequences of a more
general vierbien
ansatz, but for simplicity we remain with choice
(\ref{weproudlyuse}) (see
also  \cite{Ferraro:2011us}).

\subsection{Background $f(T)$ cosmology}

\label{fTcosm}

Let us now present the background cosmological equations in a
universe
governed by $f(T)$ gravity. Variation of the action
(\ref{action}) with
respect to the vierbein gives the equations of motion
\begin{widetext}
\begin{eqnarray}\label{eom}
e^{-1}\partial_{\mu}(eS_{A}{}^{\mu\nu})[1+f_{,T}]
-e_{A}^{\lambda}T^{\rho}{}_{\mu\lambda}S_{\rho}{}^{\nu\mu} +
S_{A}{}^{\mu\nu}\partial_{\mu}({T})f_{,TT}-\frac{1}{4}e_{A}^{\nu
}[T+f({T})]
= 4\pi Ge_{A}^{\rho}\overset {\mathbf{em}}T_{\rho}{}^{\nu},
\end{eqnarray}
\end{widetext}
where $f_{,T}$ and $f_{,TT}$  denote respectively  the
first and second derivatives of the function $f(T)$ with respect
to $T$, and
the mixed indices are used as in $S_A{}^{\mu\nu} =
e_A^{\rho}S_{\rho}{}^{\mu\nu}$. Note that the tensor
$\overset{\mathbf{em}}{T%
}_{\rho}{}^{\nu}$ on the right-hand side is the usual
energy-momentum tensor.

If we assume the background to be a perfect fluid, then the
energy momentum
tensor takes the form $\overset{\mathbf{em}}{T}_{\mu \nu
}=pg_{\mu \nu
}-(\rho +p)u_{\mu }u_{\nu }$, where $u^{\mu }$ is the fluid
four-velocity.
Under this, one sees that equations (\ref{eom}) lead to the
background
(Friedmann) equations
\begin{eqnarray}
&&H^{2}=\frac{8\pi G}{3}\rho
_{m}-\frac{f({T})}{6}-2f_{,T}H^{2}
\label{background11} \\
&&\dot{H}=-\frac{4\pi G(\rho
_{m}+p_{m})}{1+f_{,T}-12H^2f_{,TT}%
}.  \label{background2}
\end{eqnarray}%
In these expressions we have introduced the Hubble parameter
$H\equiv \dot{a}%
/a$, where a dot denotes a derivative with respect to coordinate
time $t$.
Moreover, $\rho _{m}$ and $p_{m}$ stand respectively for the
energy density
and pressure of the matter content of the universe, with
equation-of-state
parameter $w_{m}=p_{m}/\rho _{m}$. Finally, we have employed the
very useful
relation
\begin{equation}
T=-6H^{2},  \label{TH2}
\end{equation}%
which straightforwardly arises from evaluation of (\ref{telelag})
for the
unperturbed metric.

\section{The background solution of  $f(T)$ matter
bounce\label{matter bounce
background}}

In this section we examine how cosmological scenarios governed
by $f(T)$ gravity can produce a cosmological bounce. There are
two distinct points
in such an investigation. The first is to
examine whether the background evolution allows for bouncing
solutions. If this is indeed possible, then the second point is
to examine the evolution of perturbations through the bounce. The
first task is the subject of this section, while the second one
will be investigated in the next section. Finally, we mention
that in order to be closer to the convention of the literature on
this field,  we use $M_{Pl}$ instead of
$G$ when necessary, using the relation  $M_{Pl}=1/\sqrt{8\pi G}$.

In principle, whether a universe is expanding or contracting
depends on the positivity of the Hubble parameter. In the
contracting phase that exists prior to the bounce, the Hubble
parameter $H$ is negative, while in the expanding one that exists
after it we have $H>0$. By making use of the continuity equations
it follows that at the bounce point $H=0$. Finally, it is easy to
see that throughout this transition $\dot H> 0$. On the other
hand, for the transition from expansion to contraction, that is
for the cosmological turnaround, we have $H>0$ before and $H<0$
after, while exactly at the turnaround point we have $H=0$.
Throughout this transition $\dot H < 0 $.

Having in mind the above general requirements for a cosmological
bounce, and
observing the Friedmann equations (\ref{background11}),
(\ref{background2}),
we deduce that such a behavior can be easily obtained in
principle in the
context of $f(T)$ cosmology. In particular, one starts with a
specific,
desirable form of the bouncing scale factor $a(t)$, and thus one
immediately
knows $H(t)$. Concerning the matter fluid content of the
universe, with
equation-of-state parameter $w_{m}$, its evolution equation
$\dot{\rho}%
_{m}+3H(1+w_{m})\rho _{m}=0$ straightforwardly gives the solution
$\rho
_{m}(t)$ since $a(t)$ is known. Thus, we can insert these
relations in (\ref%
{background11}), and determine $f(T)$, that is $f(-6H^{2})$,
which generates
such an $H(t)$ solution. Although this procedure can always be
performed
numerically, in the following we present a simple example of a
bouncing
solution that allows for analytical results.

We start with a bouncing
scale factor of the form
\begin{equation}
 a(t)=a_{B}\left( 1+\frac{3}{2}\sigma t^{2}\right) ^{1/3},
\label{at}
\end{equation}%
where $a_{B}$ is the scale factor at the bouncing point, and
$\sigma$ is a positive parameter which describes how fast the
bounce takes place. Such an ansatz presents the bouncing
behavior, corresponding to matter-dominated
contraction and expansion, and additionally it exhibits the
advantage of allowing for semi-analytic solutions. In such a
scenario $t$ varies between $-\infty $ and $+\infty $, with $t=0$
the bounce point. Finally, in the following we normalize the
bounce scale factor $a_{B}$ to unity.

Straightforwardly we find
\begin{equation*}
H(t)=\frac{\sigma t}{(1+3\sigma t^{2}/2)}~,~~T(t)=-\frac{6\sigma
^{2}t^{2}}{%
\left( 1+\frac{3}{2}\sigma t^{2}\right) ^{2}}.
\end{equation*}%
Therefore, provided $-\sqrt{\frac{2}{3\sigma }}\leqslant
t\leqslant \sqrt{%
\frac{2}{3\sigma }}$, the inversion of this expression gives
the $t(T)$
relation as
\begin{equation}
t(T)=\pm \bigg(-\frac{4}{3T}-\frac{2}{3\sigma
}+\frac{4\sqrt{T\sigma
^{3}+\sigma ^{4}}}{3T\sigma ^{2}}\bigg)^{1/2},  \label{tinvT}
\end{equation}%
where we have kept the solution pair that gives the correct
($t=0$ at $T=0$)
behavior. Notice that when $t>$ $\sqrt{\frac{2}{3\sigma }}$ and
$t<-\sqrt{%
\frac{2}{3\sigma }}$ we have assumed the usual Einstein gravity
or the TEGR
to be the prevailing framework, thus negating the need to
pursue an $%
f\left( T\right)$ action in that region. Furthermore, we assume
the matter
content of the universe to be dust, namely $w_{m}\approx 0$.
When inserted in the evolution-equation, this leads to the usual
dust-evolution $%
\rho _{m}=\rho _{mB}a_{B}^{3}/a^{3}$, with $\rho _{mB}$ its value
at the
bouncing point.

Inserting the above expressions into (\ref{background11}) we
obtain a
differential equation for $f(t)$, which can be easily solved
analytically as
\begin{eqnarray}
f(t) &=&\frac{4t}{(2+3\sigma t^{2})M_{pl}^2}\times
\bigg[\frac{\rho _{mB}}{t}+\frac{%
6tM_{pl}^2\sigma ^{2}}{2+3t^{2}\sigma }  \notag  \label{ft} \\
&&+\sqrt{6\sigma }\rho _{mB}\,\text{ArcTan}\left(
\sqrt{\frac{3\sigma }{2}}%
t\right) \bigg]~.
\end{eqnarray}%
We mention that in the calculation we have set the integration
constant to be
zero so that the solution is consistent with the Friedmann
equation. Thus,
the corresponding $f(T)$ expression that generates a bouncing
scale factor
of the form (\ref{at}) arises from expression (\ref{ft}) with the
insertion
of the $t(T)$ relation from (\ref{tinvT}). Note that the solution
(\ref{ft}) is an
even function of $t$, and thus the $\pm$ solutions of
(\ref{tinvT})
correspond to the contraction and expansion phase respectively.
Obviously they give the same form of $f\left( T\right) $.
\begin{figure}[tbph]
\includegraphics[scale=0.31]{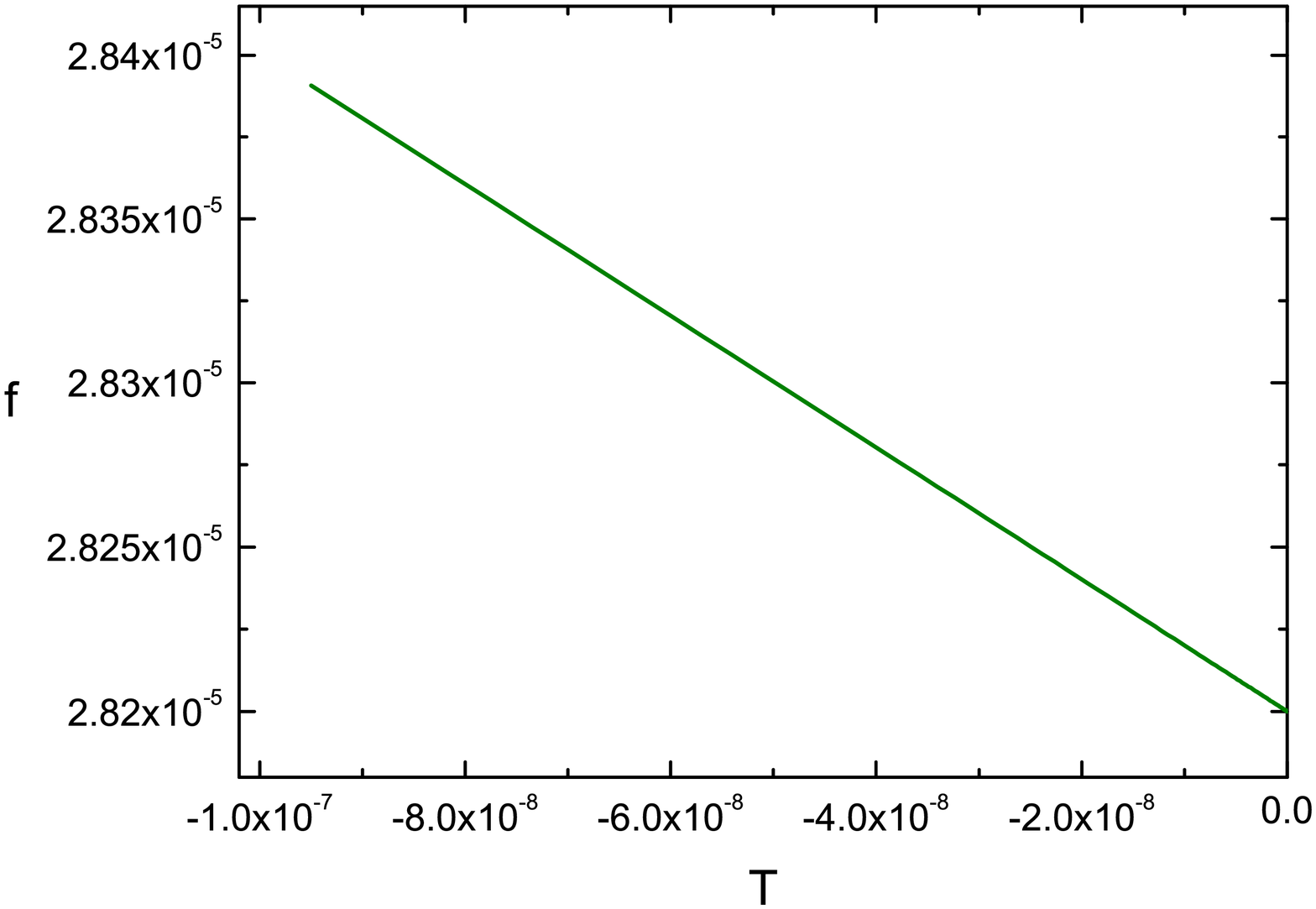}
\caption{{\it{The form of $f$ as a function of the torsion scalar
$T$ in matter bounce cosmology. The
parameters $\sigma$ and $\rho _{mB}$ were chosen to be $\sigma =
7 \times 10^{-6} M_{pl}^2$ and $\rho _{mB} = 1.41 \times
10^{-5}M_{pl}^{4}$ respectively, and the graph is in
units of $M_{pl}$.}} }
\label{Fig-fofT}
\end{figure}

In order to present this behavior more transparently, in Fig.
\ref{Fig-fofT} we depict $f(T)$ that generates the
matter-dominated bounce in $f(T)$ gravity, with $a_B=1$,
$\sigma = 7 \times 10^{-6} M_{Pl}^2$ and $\rho_{mB} = 1.41 \times
10^{-5}M_{Pl}^{4}$. {\tc {Note that, the value of $\sigma$ is
roughly determined by the amplitude of the CMB spectrum, and that
of $\rho_{mB}$ depends on how fast the standard Einstein gravity
is recovered in the $f(T)$ theory. Their physical meanings will
be discussed in the next section in detail.}}

\begin{figure}[tbph]
\includegraphics[scale=0.31]{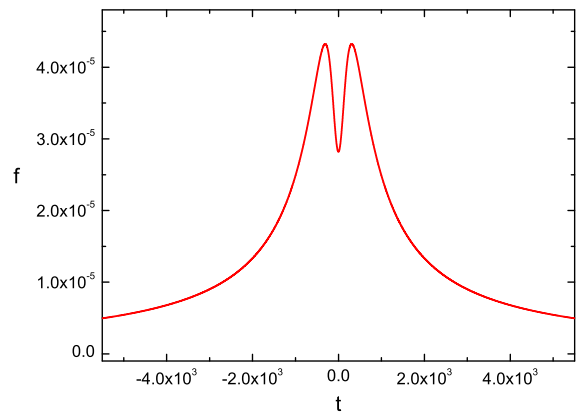}
\caption{{\it{Evolution of the function $f$ in terms of the
cosmic time $t$ in matter bounce cosmology.  The
parameters $\sigma$ and $\rho _{mB}$ were chosen to be $\sigma =
7 \times 10^{-6} M_{pl}^2$ and $\rho _{mB} = 1.41 \times
10^{-5}M_{pl}^{4}$ respectively, and the graph is in
units of $M_{pl}$.}} }
\label{Fig-ft}
\end{figure}
\begin{figure}[tbph]
\includegraphics[scale=0.31]{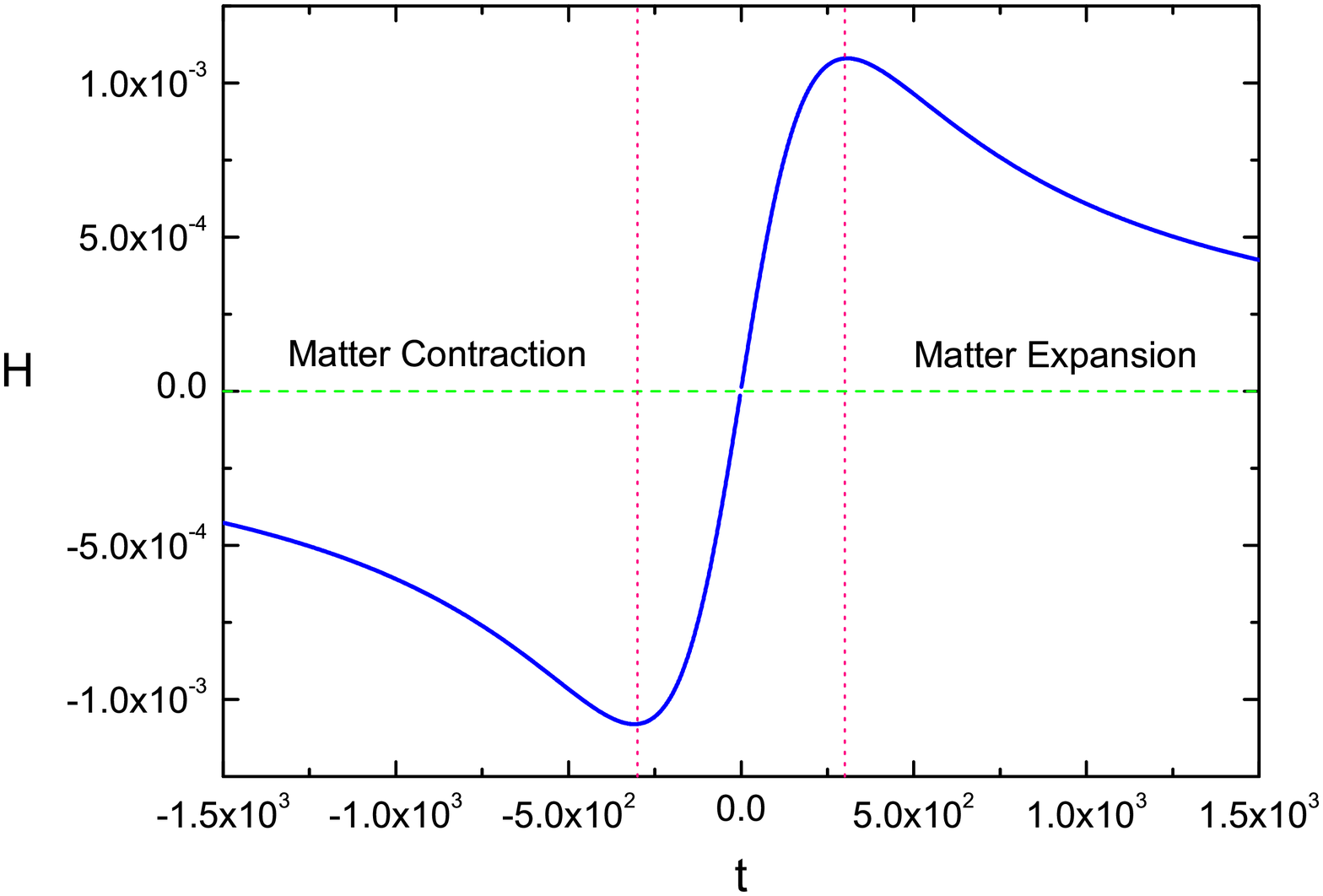}
\caption{{\it{Evolution of the Hubble parameter $H$ in terms of
the cosmic time $t$ in matter bounce cosmology.  The
parameters $\sigma$ and $\rho _{mB}$ were chosen to be $\sigma =
7\times 10^{-6} M_{pl}^2$ and $\rho _{mB} = 1.41 \times
10^{-5}M_{pl}^{4}$ respectively, and the graph is in
units of $M_{pl}$.}} }
\label{Fig-hubble}
\end{figure}
Furthermore, we numerically derived the evolution of the $f(T)$
and the Hubble parameter $H$ as functions of the cosmic time in
Figs. \ref{Fig-ft} and \ref{Fig-hubble} respectively.
Particularly, Fig. \ref{Fig-ft} shows that the evolution of
$f(T)$ is symmetric with respect to the bouncing point  $t=0$.
At the bouncing point $f$ arrives at a minimal value $16\pi
G\rho_{mB}$, which happens to cancel the contribution of normal
matter fields, and thus leads to the nonsingular bounce. From
Fig. \ref{Fig-hubble}, one can read that the background evolution
of the universe follows the usual Einstein gravity away from the
bouncing phase, but it is dominated by $f(T)$ in the middle
period. These feature are completely consistent with the designs
of the scenario as expected.

\section{Cosmological perturbations in the $f(T)$ matter bounce
\label{perturbation}}

In the previous section we presented a simple realization of the
cosmological bounce in $f(T)$ cosmology, at the background level.
In this section we extend our analysis to an investigation of the
perturbations.

We begin with a brief discussion of the cosmological evolution of
primordial perturbations in the framework of a flat FRW universe.
A standard process for generating a primordial power spectrum
suggests that cosmological fluctuations should initially emerge
inside a Hubble radius, then exit it in the primordial epoch, and
finally re-enter at late times. This process can be achieved in
the matter bounce cosmology (see for example
\cite{Starobinsky:1979ty}). In this scenario there exist quantum
fluctuations around the initial vacuum state, well in advance of
the time when the bouncing point is reached. Along with the
matter-like contraction, the quantum fluctuations would exit the
Hubble radius, since the Hubble radius decrease faster than the
wavelengths of the primordial fluctuations. When passing through
the bouncing point, the background evolution could affect the
scale dependence of the perturbations at ultra-violet scales.
However, the observable primordial perturbations, responsible for
the large scale structure of our universe, are mainly originated
in the infrared regime, where the modified gravity effect becomes
very limited  \cite{Cai:2008qw}. Consequently, in a generic case,
one can estimate the formation of primordial power spectrum with
the standard cosmological perturbation theory. In the following,
we study the perturbation theory in $f(T)$
cosmology in detail and we verify this statement in a specific
model of matter bounce cosmology.

\subsection{Generic analysis}

To begin with, we shall work in the longitudinal gauge which only involves
scalar-type metric fluctuations as
\begin{eqnarray}
ds^2=(1+2\Phi)dt^2-a^2(t)(1-2\Psi)d\bf{x}^2~,
\end{eqnarray}
thus, as usual, the scalar metric fluctuations are characterized by two
functions $\Phi$ and $\Psi$. {{Correspondingly, the fluctuation of
torsion scalar at leading order is given by
\begin{eqnarray}
 \delta{T} = 12 H (\dot\Phi+H\Psi)~,
\end{eqnarray}
which will be widely used in the following calculation.}}

By expanding the gravitational equations of motion to linear
order, we obtain the following perturbation equations
\cite{Chen:2010va} \footnote{ Note that the perturbed
gravitational equations shown in Ref.  \cite{Chen:2010va} are
incomplete due to appropriate approximations made in that paper,
but are irrelevant to the analysis. All the missing terms have
been included in the following equations.},
%
\\
\begin{widetext}
\begin{align}
\label{delta_00}
& \left(1+f_{,T}\right)\frac{\nabla^2}{a^2}\Psi
-3(1+f_{,T})H\dot\Psi -3(1+f_{,T})H^2\Phi
+36f_{,TT}H^3(\dot\Psi+H\Phi) &=& 4{\pi}G\delta\rho~, \\
\label{delta_0i}
 &(1+f_{,T}-12H^2f_{,TT})(\dot\Psi+H\Phi) &=& 4{\pi}G\delta q~,
\\
\label{delta_ij}
 &(1+f_{,T})(\Psi-\Phi) &=& 8{\pi}G\delta s~, \\
\label{delta_ii}
 &(1+f_{,T}-12H^2f_{,TT})\ddot\Psi
+H(1+f_{,T}-12H^2f_{,TT})\dot\Phi && \nonumber\\
&+3H(1+f_{,T}-12H^2f_{,TT}-12\dot{H}f_{,TT}+48H^2\dot{H}f_{,TTT}
)\dot\Psi && \nonumber\\
&+[3H^2(1+f_{,T}-12H^2f_{,TT})+2\dot{H}(1+f_{,T}-30H^2f_{,TT}
+72H^4f_{,TTT})]\Phi && \nonumber\\
& +\frac{1+f_{,T}}{2a^2}\nabla^2(\Psi-\Phi) &=& 4{\pi}G\delta p~.
\end{align}
\end{widetext}
 The functions $\delta
\rho $, $\delta {p}$, $\delta {q}$, and $\delta {s}$ are the
fluctuations of energy
density, pressure, fluid velocity, and anisotropic stress,
respectively.
We take the matter component to be a canonical scalar field $\phi
$ with a Lagrangian in form of
\begin{equation*}
\mathcal{L}=\frac{1}{2}\partial _{\mu }\phi \partial ^{\mu }\phi
-V(\phi )~,
\end{equation*}
and thus we acquire
\begin{eqnarray}
\delta \rho &=&\dot{\phi}(\delta \dot{\phi}-\dot{\phi}\Phi
)+V_{,\phi
}\delta \phi ~,  \label{delta_p} \\
\delta q &=&\dot{\phi}\delta \phi ~,  \label{delta_q2} \\
\delta s &=&0~,  \label{delta_s1} \\
\delta p &=&\dot{\phi}(\delta \dot{\phi}-\dot{\phi}\Phi
)-V_{,\phi }\delta
\phi ~,
\end{eqnarray}
respectively.

Inserting relation (\ref{delta_s1}) into  (\ref{delta_ij})
one can obtain $%
\Psi =\Phi $ due to a vanishing anisotropic stress, which is also
widely
found in the standard theory of cosmological perturbations (for
example see
 \cite{Mukhanov:1990me}). Moreover, combining
(\ref{delta_0i}) and (\ref{delta_q2}) implies that the
gravitational potential $\Phi $ can be completely determined by
the scalar
field fluctuation $\delta \phi $. Therefore we conclude that for
our choice of the tetrad given in (\ref{weproudlyuse}), there
exists
only a single degree of freedom in the scenario of $f(T)$
gravity minimally
coupled to a canonical scalar field. Note that there is another
evolution equation to describe the dynamics of cosmological
perturbations, namely the
perturbed equation of motion for $\delta \phi $. However, using
(\ref{delta_0i}) it can be shown that this second equation is
consistent with (\ref{delta_00}) and (\ref{delta_ii}).

In order to understand the evolution of scalar-sector metric
perturbations, we
use the perturbed equation of motion for the gravitational
potential $\Phi$ instead of the scalar field fluctuation
$\delta\phi$. Combining  (\ref{delta_00}),
(\ref{delta_ii}) and (\ref{delta_0i}) we
obtain the complete form of the equation of motion of one Fourier
mode $%
\Phi_k$ as:
\begin{eqnarray}  \label{eom_Phi_com}
 \ddot\Phi_k + \alpha \dot\Phi_k + \mu^2 \Phi_k + c_s^2
\frac{k^2}{a^2}\Phi_k = 0~,
\end{eqnarray}
with
\begin{eqnarray}
 \alpha &=& 7H+\frac{2V_{,\phi}}{\dot\phi} -\frac{36H\dot{H}
 (f_{,TT}-4H^2f_{,TTT})}{1+f_{,T}-12H^2f_{,TT}}~, \\
 \mu^2 &=& 6H^2 +2\dot{H} +\frac{2HV_{,\phi}}{\dot\phi}
\nonumber \\
 &&
-\frac{36H^2\dot{H}(f_{,TT}-4H^2f_{,TTT})}{1+f_{,T}-12H^2f_{,TT}}
~, \\
 c_s^2 &=& \frac{1+f_{,T}}{1+f_{,T}-12H^2f_{,TT}}~,
\end{eqnarray}
where we have applied the relation $\Psi=\Phi$ holding in
the absence of anisotropic
stress. The functions $\alpha$, $\mu^2$ and $c_s^2$ are
respectively the frictional
term, the effective mass, and the sound speed parameter for the
gravitational potential $\Phi$. Moreover,
we recall the
scalar-field background equation
\begin{eqnarray}
 \ddot\phi + 3H\dot\phi + V_{,\phi} = 0,
\end{eqnarray}
and the second Friedmann equation (\ref{background2}), which in
our case reads:
\begin{eqnarray}
 (1+f_{,T}-12H^2f_{,TT})\dot{H} = -4\pi G\dot\phi^2~.
\end{eqnarray}
Consequently, to make use of these two background equations,
(\ref{eom_Phi_com}) can be further simplified as
\begin{equation}  \label{eom_Phi_sim}
 \ddot\Phi_k +\left(H-\frac{\ddot{H}}{\dot{H}}\right)\dot\Phi_k +
\left(2\dot{H}-\frac{H\ddot{H}}{\dot{H}}\right)\Phi_k +
\frac{c_s^2 k^2}{a^2}\Phi_k = 0 .
\end{equation}
Surprisingly, we find that the equation of motion for the
gravitational
potential in the present $f(T)$ scenario is the same with that in
the standard Einstein
gravity \cite{Mukhanov:1990me}, except the newly introduced sound
speed
parameter. This important feature could be a key for us to
explore potential
clues of the $f(T)$ theory in cosmological surveys.

\subsection{Variables of perturbations in bouncing cosmology}

After having derived the equation of motion for the gravitational
potential, we proceed in solving it in the
detailed bouncing cosmology. As we showed in the background
analysis of section \ref{matter bounce
background}, we can achieve the exact matter bounce
scenario in which the universe evolves as a matter-dominated one
in the
contracting phase. In order to accommodate this picture, the
matter
component could be a massive scalar field similar to the
mechanism of the
Lee-Wick bounce \cite{Cai:2008qw}, or a scalar field with a
fine-tuned
exponential potential \cite{Wands:1998yp}.

One often uses a gauge-invariant variable $\zeta$, the curvature
fluctuation in comoving coordinates, to characterize the cosmological
inhomogeneities. In the case of $f(T)$ cosmology we assume that the form of
$\zeta$ is the same as that defined in the standard cosmological
perturbation theory, which is given by
\begin{eqnarray}  \label{zeta}
\zeta = \Phi-\frac{H}{\dot H}\left(\dot\Phi+H\Phi\right)~.
\end{eqnarray}
A useful relation for the time derivative of $\zeta$ can be derived upon
making use of equation (\ref{eom_Phi_sim}), namely
\begin{eqnarray}  \label{dot_zeta}
\dot\zeta_k = \frac{H}{\dot{H}}\frac{c_s^2k^2}{a^2}\Phi_k~.
\end{eqnarray}
In a generic expanding universe $\dot\zeta_k$ approaches zero at large
length scales, $k\rightarrow0$, since the dominant mode of $\Phi_k$ is then
nearly constant. However, in the matter bounce cosmology the metric
perturbation $\Phi_k$ in the contracting phase is dominated by a growing
mode with $\Phi_k\sim k^{-7/2}$, and thus $\zeta$ keeps increasing before
arriving at the bouncing point \cite{Cai:2008ed}.

{{Note that, one may be concerned of the variable $\zeta$ becomes
ill-defined when $\dot H$ changes sign. In fact, this specious trouble was
extensively studied in many aspects of cosmological perturbation theory. At
present, our understanding on a well-defined cosmological perturbation
theory is to require the metric perturbation and the corresponding
extrinsic curvature behave smoothly throughout the background evolution.
The pioneer discussion on this topic appeared in Ref. \cite{Wands:1998yp},
and we refer to Ref. \cite{Cai:2008qw} for recent detailed analysis by
tracking the evolution of metric perturbation step by step in matter bounce
cosmology. In our calculation we still make use of $\zeta$ merely since its
analytic analysis is very convenient to be performed away from the bouncing
phase. In addition, we would like to point out that the knowledge
obtained in General Relativity can be also applied to $f(T)$
gravity, which is a modification of Einstein gravity, in the frame of
bouncing cosmology. The reason is that the observable modes of
perturbations in bouncing cosmology are distributed in the Infrared regime
and thus the effect caused by the modification of Einstein gravity has to
be very limited. }}

In order to simplify the calculation, we further introduce a canonical
variable to characterize the cosmological perturbations
\begin{eqnarray}  \label{v_var}
v = z\zeta~,
\end{eqnarray}
where
\begin{eqnarray}  \label{z_var}
z \equiv a\sqrt{2\epsilon}~,
\end{eqnarray}
with $\epsilon\equiv-\frac{\dot{H}}{H^2}$. It can then be found
that the
equation of motion
\begin{eqnarray}  \label{v_eom_g}
v_k^{\prime \prime }+(c_s^2k^2-\frac{z^{\prime \prime
}}{z})v_k=0~,
\end{eqnarray}
where the prime denotes the derivative with respect to the
comoving time $%
\tau\equiv\int dt/a$,
 is still available in a contracting
universe.

\subsection{The primordial power spectrum of the $f(T)$ matter
bounce}

In order to perform a specific analysis we recall that in the
matter-like
contracting phase the scalar factor evolves as
\begin{eqnarray}
a \sim t^{2/3} \sim \tau^2~,
\end{eqnarray}
and $z\propto a$. In this period, by solving the background
equations of
motion, one obtains the approximate relations
\begin{eqnarray}\label{fTmatter}
H\simeq\frac{2}{3t}~,~
f(T)\simeq\left(-1+\frac{\rho_{mB}}{2M_{pl}^2\sigma}%
\right)T~,
\end{eqnarray}
that
hold far  before the beginning of the bouncing era
$t_m=-\sqrt{%
\frac{2}{3\sigma}}$. As we mentioned above,
$\rho_{mB}$ is the energy density of the
matter field at the bouncing point and $\sigma$ describes how
fast the
bounce takes place. Therefore, the sound speed of the curvature
perturbation
reverts to $c_s^2\simeq1$ in the matter-like contracting phase.

We mention here that inserting the $f(T)$ form of
(\ref{fTmatter}) into the action (\ref{action}), we find
that the standard Einstein gravity will automatically be
recovered when $\rho_{mB} \simeq 2M_{Pl}^2\sigma$. Particularly,
when $\rho_{mB}$ exactly equals to $2M_{Pl}^2\sigma$, we get
$f(T)\sim O(T^2)$, which will dilute faster than the Ricci scalar
during late-time evolution. Even when $\rho_{mB}$ is not equal to
$2M_{Pl}^2\sigma$, it is clear that the system satisfies General
Relativity with a rescaled gravitational constant. Thus, the
combination of $\rho_{mB}$ and $\sigma$ could, in principle, be
constrained by measurements of the gravitational constant. In our
computation we choose $\rho_{mB}$ to be slightly different
from $2M_{Pl}^2\sigma$, thus our model is able to
approach the standard Einstein theory far away from the bounce.

As a consequence, the perturbation equation becomes
\begin{equation}
v_{k}^{\prime \prime }+\left( k^{2}-\frac{2}{\tau ^{2}}\right)
v_{k}\simeq
0~,  \label{v_eom_m}
\end{equation}%
in the contracting phase. Initially the $k^{2}$-term dominates
in (\ref%
{v_eom_m}) and thus we can neglect the gravitational term. From
this point
of view, the fluctuation corresponds to a free scalar propagating
in a flat
spacetime, and naturally the initial condition takes the form of
the
Bunch-Davies vacuum \footnote{%
As discussed in the previous subsections, the generic form of the
perturbation
equation involves a sound speed parameter, and correspondingly
the
Bunch-Davies vacuum initial condition should take the form
$v_{k}\simeq {%
e^{-ik\tau }}/\sqrt{2c_{s}k}$.}:
\begin{equation*}
v_{k}\simeq \frac{e^{-ik\tau }}{\sqrt{2k}}~.
\end{equation*}%
Making use of the vacuum initial condition we solve the
perturbation
equation exactly, obtaining the solution
\begin{equation*}
v_{k}=\frac{e^{-ik\tau }}{\sqrt{2k}}\bigg(1-\frac{i}{k\tau
}\bigg)~.
\end{equation*}%
From this result we deduce that the quantum fluctuations
could become
classical perturbations, after exiting the Hubble radius, due to
the
gravitational term in equation (\ref{v_eom_m}). Moreover, the
amplitude of the
metric perturbations will keep increasing until the universe
arrives at the
bouncing phase at the moment $t_{m}$.

From the definition of the power spectrum we see that
$\zeta\sim
k^{3/2}|v_k|$ is scale-invariant in our model, which can also be
achieved in
inflationary cosmology. However, the coefficient $\epsilon$ takes
the value $%
\frac{3}{2}$ in the matter-like contraction and thus it is unable
to amplify
the power spectrum of metric perturbation as in inflation.
A detailed
calculation provides the expression of the primordial power
spectrum for the $%
f(T)$ matter bounce as
\begin{eqnarray}
P_{\zeta} \equiv \frac{k^3}{2\pi^2}\left|\frac{v_k}{z}\right|^2 =
\frac{H_m^2}{%
48\pi^2M_{Pl}^2}~,
\end{eqnarray}
where $H_m=\sqrt{{\sigma}/{6}}$ is the absolute value of the
Hubble
parameter at the beginning moment of the bouncing phase.  It should be
noted that the $M_{Pl}^2$ which appears in the power spectrum will become
rescaled when $\rho_{mB}$ is not equal to $2M_{Pl}^2\sigma$.

\subsection{The tensor-to-scalar ratio and the matter
bounce
curvaton scenario}

We can now study the amplitude of tensor perturbations $h_{ij}$
in the $%
f(T)$ matter bounce. Following  \cite{Chen:2010va}, the
perturbation
equation for the tensor modes can be expressed as
\begin{equation}
\bigg(\ddot{h}_{ij}+3H\dot{h}_{ij}-\frac{\nabla
^{2}}{a^{2}}h_{ij}\bigg)-%
\frac{12H\dot{H}f_{,TT}}{1+f_{,T}}\dot{h}_{ij}=0~,  \label{h_eom}
\end{equation}%
while the tensor modes are transverse and traceless, namely
\begin{equation*}
\partial _{i}h^{ij}=h_{i}^{i}=0~.
\end{equation*}

In the specific case of matter bounce cosmology, equation
(\ref{h_eom}) is the
same as the standard one since $f_{,TT}$ vanishes in the
contracting phase.
Hence, following the analysis done in \cite{Cai:2008qw},
the primordial
power spectrum of tensor fluctuations is also scale-invariant,
but the
amplitude is $h\sim H_m/M_{Pl}$, which is of the same order of
the scalar
perturbation. Therefore, the $f(T)$ matter bounce scenario
suffers from the usual
problem of all matter bounce models, namely that the ratio
of tensor to scalar $r\equiv
P_T/P_{\zeta}$ is difficult to accommodate with current
observations. In particular, from the current CMB data
\cite{Komatsu:2010fb} this
ratio is required to be less than $0.2$.

Consequently, in order to make our model consistent with
cosmological observations on the value of $r$, it is necessary to
introduce a mechanism
to magnify the amplitude of scalar-type metric perturbations.
This
issue can be resolved by introducing additional light scalar
fields, as in the
matter bounce curvaton scenario \cite{Cai:2011zx}. These scalars
are able to
seed isocurvature fluctuations, and then transfer to a
scale-invariant
spectrum of the adiabatic fluctuations during the nonsingular
bouncing phase,
through the so-called kinetic amplification. Thus, we
obtain a mechanism for
enhancing the primordial adiabatic fluctuations and suppressing
the tensor-to-scalar ratio.

Following the analysis performed in \cite{Cai:2011zx} we
can introduce
one massless scalar field $\chi$, coupling to the background
matter field
through the term $g^2\phi^2\chi^2$. Therefore, the
tensor-to-scalar ratio in
our model can be suppressed by the kinetic amplification factor
in the
bouncing phase, which can be expressed as $r\simeq
\mathcal{F}^{-2}$ as
shown in equation (58) of \cite{Cai:2011zx}. Specifically,
in our case the
kinetic amplification factor $\mathcal{F}$ is determined to be
$\mathcal{F}%
\simeq18.66$, and so $r\simeq2.87\times10^{-3}$. Finally, by
virtue of the
matter bounce curvaton mechanism, our model is able to satisfy
the
constraints from current observations.

\section{Conclusions and Discussion\label{conclusion}}

In this work we investigated the realization of matter bounce cosmology in
the framework of $f(T)$ gravity. Considering an explicit scale-factor form,
which links the matter-like contraction and matter-like expansion through a
smooth nonsingular bouncing phase, we reconstructed the specific form of
$f(T)$ that generates it. Our analysis has illustrated the possibility of
combining the $f(T)$ gravity with bounce cosmologies, and thus exploring an
alternative approach to avoid the Big Bang singularity encountered in the
standard inflationary cosmology. Additionally, the constructed $f(T)$
matter bounce  is free of ghost degrees of freedom and other
potential problems of many matter bounce scenarios, which is a significant
advantage.

{{At the background level, we found that the theory of torsion
gravity is difficult to be verified by experiments. This is in agreement
with the mechanism described in  \cite{Buchbinder:1985ym}, where the
torsion dynamics is highly degenerated with quantum corrections to the
classical gravitational action caused by vacuum effects.}} However, going
beyond the background level, we performed a detailed analysis of the
evolution of metric perturbations in the present work. For our choice of
the tetrad we found that the scalar-type metric perturbations posses a
single degree of freedom, but they have a time-dependent sound speed, a
feature that is generic in any $f(T)$ scenario. We expect that this
behavior could be an important signature to be detected or constrained by
observations.

The present $f(T)$ matter bounce scenario suffers from the usual problem,
widely existing in many other bounce models, namely it predicts a value for
tensor-to-scalar ratio larger than observational constraints
\cite{Cai:2008qw, Wands:2008tv}. This undesirable feature can be cured by
the mechanism of matter bounce curvaton, which introduces an additional
scalar field tracking the background evolution before the bounce, but with
fluctuations obtaining a kinetic amplification during the bouncing phase.

Finally, we would like to point out that in order to establish the
existence of matter bounce in $f(T)$ gravity and perform a first
examination of its basic features, we constructed a simplified scenario,
without the incorporation of the contribution of radiation, dark energy and
other matter components. Therefore, the investigated scenario involves only
two parameters, namely the energy density at the bouncing point $\rho_{mB}$
and the bouncing parameter $\sigma$. Through the perturbation analysis we
found that these two parameters could be tightly constrained by CMB
observations and experimental bounds on the Newton's constant. It would be
interesting to construct bouncing models in $f(T)$ gravity going beyond
this scenario, in which the contributions of other components would be
taken into account and various cosmological observations would be
satisfied. We leave such a construction of a phenomenologically realistic
model and its observational constraints for a future investigation.

\begin{acknowledgments}
We wish to thank Damien A. Easson and Tanmay Vachaspati for stimulating
discussions. The work of Y.F.C., S.H.C, J.B.D, and S.D. is supported in
part by funds of Physics Department at Arizona State University.
\end{acknowledgments}

\end{document}